\documentclass[twocolumn,showpacs,superscriptaddress,preprintnumbers,amsmath,amssymb,epsfig,color]{revtex4}

\usepackage{dcolumn}
\usepackage{bm}

\usepackage{graphicx}

\newcommand\be{\begin{equation}}
\newcommand\ee{\end{equation}}
\newcommand\ba{\begin{eqnarray}}
\newcommand\ea{\end{eqnarray}}
\newcommand\eq{\begin{equation}}           
\newcommand\en{\end{equation}}

\input epsf
\usepackage{amssymb}
\include{epsf,epsfig}

\usepackage{subfigure}

\begin{document}
\title{
Positrons in Cosmic Rays from Dark Matter Annihilations\\
 for Uplifted Higgs Regions in MSSM}
\author{Kenji Kadota$^1$, Katherine Freese$^1$, and Paolo Gondolo$^2$   \\
{\em \small 
$^1$ Michigan Center for Theoretical Physics, University of Michigan, Ann Arbor, MI 48109} \\
 {\em \small
$^2$ Department of Physics, University of Utah, Salt Lake City, UT 84112}
}
\begin{abstract}
We point out that there are regions in the MSSM parameter space which successfully
provide a dark matter (DM) annihilation explanation for observed 
$e^+$ excess (e.g. PAMELA), while still remaining in agreement with all other data sets. Such regions
(e.g.  the uplifted Higgs region) can realize an enhanced neutralino DM 
annihilation dominantly into leptons via a Breit-Wigner resonance through the CP-odd Higgs channel. Such regions can give
the proper thermal relic DM abundance, and the DM annihilation products
are compatible with  current antiproton and gamma ray observations.  
This scenario can succeed without introducing any additional degrees of freedom beyond those already in the MSSM.  
\end{abstract}
\maketitle   
\setcounter{footnote}{0} 
\setcounter{page}{1}
\setcounter{section}{0} \setcounter{subsection}{0}
\setcounter{subsubsection}{0}

\section{Introduction}

There has been growing interest in the interpretation of an observed positron excess 
 in light of the 
recent data from PAMELA \cite{adri1,adri2,pamelapbar} and Fermi-LAT \cite{fermi1,port1,atw},
while satisfying antiproton and gamma ray constraints.

Among the possible sources for the positron excess are  astrophysical sources, such as pulsars \cite{dan09,yuk,profu,maly,maly2,dela10} and supernovae \cite{shav,sn1,fuji,subir1,subir}, and  dark matter \cite{bal3,gordy1,bal5,cmssm3,dan4,cir,cire5,gordy3,chen5,yasu,ishi,ibarra2,nojiri2,gog,meade,moro2,gao1,eds,mea,gra,paolo5,joe,boem3,chen3,dib,mssm}. The dark matter (DM) possibility is of great interest from the particle theory viewpoint, and we seek in this paper the cosmic and gamma ray signatures of DM annihilation in a supersymmetric model within the MSSM (Minimal Supersymmetric Standard Model) parameter space.

 Most of the conventionally explored MSSM 
parameter space cannot successfully explain the observed positron data.
There is typically a significant branching ratio of DM 
 annihilation into gauge bosons and Higgs as well as quarks and hence 
too much hadronic production of antiprotons, in excess of what is observed. 
The MSSM, however, has more than a hundred 
parameters and it would be worth seeking such a possibility to realize the current cosmic ray observations in the framework of the MSSM 
without introducing any additional degrees of freedom.
Another persistent problem in the dark matter annihilation scenarios to explain the observed positron excess is the requirement of  large annihilation cross sections in the halo far bigger (typically a boost factor of a factor 100 or more) than 
the canonical thermally averaged cross section value at 
 freeze-out for the weakscale dark matter inferred from the 
observed dark matter relic density.  Several resolutions
to this apparent discrepancy in the annihilation cross sections
have been proposed, such as unconventional cosmological histories \cite{paolo4,pall,cate} which can affect the dark matter freeze-out temperature; Sommerfeld enhancement \cite{hisa1,cire,nima1} which requires  new light particles to allow new long range interactions 
between dark matter particles; Breit-Wigner enhancement \cite{posp,dan,ibe,guo} which 
requires a particle whose mass is close to the twice of the dark matter mass; and  substructure clumps \cite{boost2, die2,kuh,alb,brun,mark1} which could provide a partial contribution to 
the enhancement (say, by a factor of a few). The annihilation enhancements make the the 
current annihilation cross section higher in galaxies today than it was at the time of freeze-out, so that 
it can explain the positron excess (which requires a high cross section today)
while still satisfying the correct relic density (which requires a lower cross section at freeze-out).

The main results of the paper are to point out the existence of  parameter regions in the MSSM which can potentially realize a DM annihilation scenario that explains the observed
positron excess.  We show that there can be dominant leptonic final states 
and a large boost factor to obtain the positron excess and thermal relic abundance without introducing any additional degrees of freedom beyond those already in the MSSM; in addition
these scenarios can be compatible with the current antiproton and gamma ray data.

As a concrete example, we consider the following scenario:  The requirement of a dominant leptonic final state (in this case taus) can be satisfied in the
 uplifted Higgs regions \cite{bog} within the MSSM as described in Section \ref{mot}.   In addition, we obtain the required boost factor from
the pseudo-scalar Higgs s-channel resonance in the MSSM which can induce the Breit-Wigner enhancement \cite{posp,dan,ibe,guo} via $\chi \chi \rightarrow A \rightarrow \tau^+ \tau^-$ for $m_A \sim 2 m_\chi$
($\chi$ denotes the neutralino dark matter and $A$ denotes the CP-odd Higgs).

We also note that we have here found a situation where annihilation to taus satisfies
 all existing observations.  The reason is that we treat the FERMI data as astrophysical background (fit by a simple power law)
 while we take the PAMELA excess to be due to DM annihilation.  Our interpretation is in contrast to some of the previous literature, where 
the requirement was made of explaining the excesses in both PAMELA and FERMI as due to DM annihilation.
 
In this paper we take the dark matter to have a cored isothermal density profile \cite{bah,core} in the Galactic Halo unless stated otherwise. 
This relatively flat core density distribution  helps to ameliorate the severe gamma ray constraints \cite{cirel,pap,kev2} (from observations e.g. by FERMI and HESS)
in contrast to other profiles such as the NFW profile \cite{nfw}.   Our choice of profile should be adequate for 
the purpose of illustrating the potential significance of the previously unexplored MSSM parameter space discussed here.

Sec. \ref{mot} reviews the uplifted supersymmetric Higgs regions in the MSSM parameter space and its unique properties well motivated for the current cosmic ray observations. Sec. \ref{cosray} then discusses the cosmic and gamma rays signals expected for such regions, followed by the conclusion/discussion in Sec. \ref{conc}.

\section{Motivation: Uplifted Higgs region}
\label{mot}

The purpose of this section is to briefly review the uplifted Higgs regions (heavy Higgs mass is `uplifted' in those regions) and point out that there exists a region in the MSSM parameter space where 
the dark matter can annihilate dominantly to leptons and its annihilation cross 
section can be boosted via the Breit-Wigner resonance without introducing any additional fields beyond those 
already in the MSSM. We refer the readers to Ref. \cite{bog} for more detailed discussions on the uplifted Higgs regions.

In the usual MSSM, it is difficult to obtain the required leptophilic cross sections to explain PAMELA while not overproducing $\bar p$.
In particular, in this paper we focus on the case where the dominant channel for DM annihilation 
takes place via Higgs resonance $\chi \chi \rightarrow A \rightarrow \tau^+ \tau^-, b \bar{b}$.
In the standard MSSM, for those Higgs funnel regions with a large $\tan \beta$, the problem is that the branching ratio to $\tau ^+ \tau^-$ is typically at the 10\% level while the remainder is predominantly  to $b \bar b$; the latter
produces far too many $\bar p$ in excess of what is observed.  

  In the standard case the ratio of final state $ \tau^+ \tau^-$ to final state $b \bar b$ must be small because
  $BR (\chi \chi \rightarrow A \rightarrow \tau^+ \tau^-)/ 
 BR (\chi \chi \rightarrow A \rightarrow b \bar b) \sim y_\tau^2 / 3 y_b^2 $ and the Yukawa couplings
  $y_b > y_{\tau}$ since $m_b(=y_b \langle H_d \rangle) > m_{\tau}(=y_\tau \langle H_d \rangle)$.  
  In the uplifted Higgs regions, on the other hand, the bottom type fermion mass and the bottom type Yukawa coupling are not necessarily proportional to each other, so that one can have $y_\tau > y_b$ and the dominant annihilation to $\tau$.  Unlike the standard case,
  here  $m_\tau$ is generated not by the usual tree level down-type Higgs vacuum expectation value (VEV) $\langle H_d \rangle$
 (which vanishes at tree level)
 but instead at the one loop level from $\langle H_u \rangle$ and $\langle H_d \rangle$.

Such unique features in the uplifted Higgs regions arise by the absence of the soft-SUSY breaking B-term, $B \mu H_u H_d$ (which can be 
justified for instance by the appropriate R-charge assignment $R(H_d,Q,U^c,E^c)=0$ and $R(H_u,D^c,L)=2$) while the superpotential  is same as that of the conventional R-parity conserving MSSM. 
This results in the vanishing down-type Higgs VEV at the tree level by enforcing the electroweak symmetry breaking 
via $|\mu|^2+m_{H_u}^2<0,|\mu|^2+m_{H_d}^2>0$ and the stability of $H_u H_d$ flat direction via $2|\mu|^2+m_{H_u}^2+m_{H_d}^2>0$. 
$\langle H_d \rangle \ne 0$ comes from the loop 
contributions and it helps in explaining the small down type masses which are loop suppressed. 
The down type quark and lepton masses arise at the one-loop level\ba
m_{d}=y_{d} \langle H_d \rangle +y_{d}' \langle H_u \rangle
\ea
$y_d$ is the standard down type Yukawa coupling between $H_d$ and down type fermions 
and $y_d'$ represents the 
loop-induced effective Yukawa coupling between $H_u$ and down type fermions \cite{bog}. Hence $y_d$ is not 
in general proportional to $m_d$ in the uplifted Higgs regions anymore, and, for instance, $y_{\tau} > y_b$ can be possible even though $m_{\tau} < m_{b}$.

 In particular the decay of a heavy Higgs  
into the down type fermion pairs is proportional to $y_d$ but not to $m_d \tan \beta$ in contrast to the usual MSSM. The parameter space we focus on has
 $m_A \approx 2 m_{\chi}$ and $\tan \beta \gg 1$ as well as $y_{\tau} > y_b$, so that, as discussed above,
 we consider the dominant dark matter annihilations through Higgs resonance $\chi \chi \rightarrow A \rightarrow \tau^+ \tau^-, b \bar{b}$ (note $A$ is dominated by $H_d$ component in our scenarios). The coupling of $A$ to the top quarks are highly suppressed by the factor $\cot ^2 \beta$ and the final states are dominated by $\tau$ and $b$ with an aforementioned ratio 
$Br(\chi \chi \rightarrow \tau^+ \tau^-) / Br(\chi \chi \rightarrow \bar{b}b) \sim y_{\tau}^2/(3 y_b^2)$ in our scenarios.    
Such regions have been missed in the conventional studies of MSSM, and we argue the 
potential significance of such regions for its unique cosmic ray signals distinguishable from the conventional MSSM cosmic ray predictions.

We in the following briefly discuss the enhancement of the annihilation cross sections in 
the uplifted Higgs regions to see the range of the parameters required for a sufficient boost factor . 
We simply assume the dark matter is a neutralino LSP (lightest supersymmetric particle) and we are interested in the case where its mass is close to the twice of 
the pseudo-scalar Higgs to obtain a large boost factor via the Higgs resonance. 
The relevant couplings in the Lagrangian are 
\ba
L \ni C^{ddA} \bar{d} \gamma_5 d A
 +\frac{1}{2}C^{\chi \chi A} 
\bar{\chi} \gamma_5 \chi A
\ea
with
\ba
C^{ddA}&=&i \frac{y_{d}}{\sqrt2} \sin \beta \nonumber \\ 
~C^{\chi \chi A}&=& -i (g N_{12}-g' N_{11}) (\sin \beta N_{13}-\cos \beta N_{14})
\ea
where $N_{ij}$ is the neutralino mixing matrix in the $(\tilde{B},\tilde{W}^0,\tilde{H}^0_d,\tilde{H}^0_u)$ basis, $g',g$ are $U(1),SU(2)$ gauge couplings respectively and our scenarios are dominated by $d=\tau,b$ final states. The dark matter annihilation cross section through the Higgs resonance $\sigma_{\chi \chi}(\chi\chi \to A \to d \bar{d} )$ becomes 
\ba
\sigma_{\chi \chi}= 
\frac{ |C_{ddA}|^2 |C_{\chi\chi A}|^2}{32\pi}  \frac{k} {p} \frac{s}{(s-m_A^2)^2 + \Gamma_A^2 m_A^2
}
\ea
where $s$ is the Mandelstam variable and $p= \sqrt{s-4m_\chi^2}/2$ represents the initial momentum of the incoming particle $\chi$ in the center-of-mass (c.m.) frame, $k= \sqrt{s-4m_f^2}/2$ is the final c.m. momentum and $\Gamma_A$ is the total decay width of $A$. The branching ratio $Br(\chi \chi \rightarrow b \bar{b})\sim {\mathcal O}(0.1)$ with $Br(\chi \chi  \rightarrow \tau^+ \tau^-)+BR(\chi \chi  \rightarrow b \bar{b}) \approx 1$ can be possible in the uplifted Higgs regions in contrast to the usual Higgs resonance in MSSM with a large $\tan \beta$ where $Br(\chi \chi \rightarrow \tau^+ \tau^-)\sim {\mathcal O}(0.1)$. Its thermal average is \cite{paolo3,paolo4222}
\ba
\label{rel}
\langle \sigma v\rangle &=& \frac{1}{8m^4_{\chi}TK^2_2(m_{\chi}/T)} \nonumber \\
&& \int ^{\infty}_{4 m^2_{\chi}} ds \sigma_{\chi \chi} (s)(s-4 m_{\chi}^2)
\sqrt{s}K_1(\sqrt{s}/T)    \nonumber \\ 
 &=&\frac{4x }{K_2^2(x)}
\int ^{\infty}_{0} dz \sigma(z)z \sqrt{1+z} K_1(2x\sqrt{1+z})
\ea
where we take $s=4(1+z)m_{\chi}^2$, $x\equiv m_{\chi}/T$ and 
\ba
\sigma(z)\propto \frac{(1+z)^{3/2}}{\sqrt{z}}
\frac{1}{(z+\delta)^2+\gamma^2 (1-\delta)^2}
\ea
with 
\ba
\label{defns}
m_A^2=4m_{\chi}^2(1-\delta),\gamma=\Gamma_A/m_A .
\ea
\begin{figure}[htb!]
\begin{center}    
\epsfxsize = 0.48\textwidth
\epsffile{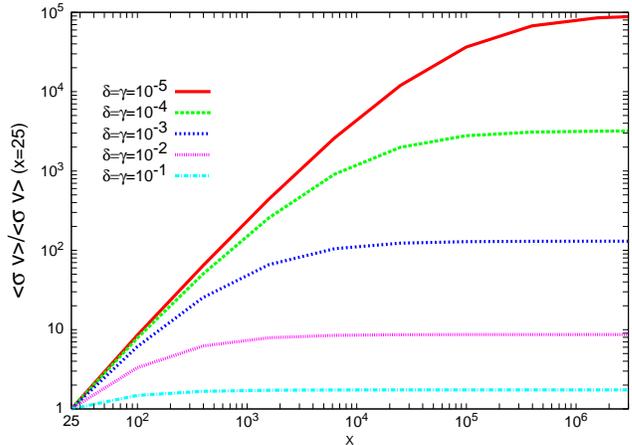}
\end{center}        
\caption{Boosted cross section:
The numerically integrated Breit-Wigner enhancement $\langle \sigma v \rangle(x)/\langle \sigma v \rangle(x_f=25)$ for, from 
the top to bottom curves, $\delta=\gamma=10^{-5},10^{-4},10^{-3},10^{-2},10^{-1}$ as
defined in Eq.(\ref{defns}). $\langle \sigma v \rangle(x=25)$ represents the usual thermally averaged annihilation cross section at `freeze out' $x\equiv m_\chi/T=25$.}
\label{avesigmav}
\end{figure}
The conventional calculation of the neutralino annihilations in terms of the 
series expansions of $x$ is 
inadequate when $m_{A}\sim 2 m_{\chi}$ and Figure \ref{avesigmav} shows the 
numerical integration of $\langle \sigma v \rangle$ normalized by its value 
at the freeze-out temperature $x_f$ (we set $x_f=25$ for illustration) 
for a few parameter sets $(\delta,\gamma)$.

Compared to the non-resonant case with  constant 
$\langle \sigma v\rangle $ after the freeze-out, the dark matter 
abundance with the Breit-Wigner resonance is enhanced by the 
effective boost factor (BF) \cite{kt} 
\ba
\label{eq:boost}
BF\equiv \frac{Y_{\infty, res}}{Y_{\infty, non-res}}
\simeq
\frac
{\langle \sigma v\rangle|_{T=0} /x_f }
{  \int _{x_f}^{\infty} dx \frac{\langle \sigma v\rangle }{x^2}  }
\ea
where we took the same cross section at
zero temperature to be $\langle \sigma v\rangle |_{T=0}$. The present relic abundance hence is
\ba
\Omega_{\chi}h^2 \simeq
0.1 \times \frac{10^{-9} GeV^{-2}}{\langle \sigma v\rangle|_{T=0}}
\frac{x_f}{\sqrt{g_*}}
\times BF
\ea
where $g_*$ is the effective number of degrees of freedom. 

We compute values for the boost factor using Eqn.(\ref{eq:boost}). For instance, 
the choice of $\delta$ $=$ $\gamma$ $=$ $10^{-1}$ $(10^{-2}$,$10^{-3}$,$10^{-4}$,$10^{-5})$ gives respectively the boost factor 
of the order a few (${\cal O}$(10), ${\cal O}$(50), ${\cal O}$(300), ${\cal O}$(3000)).
The justification of these small $\delta,\gamma$ values is beyond the scope of this paper.
Nonetheless, we consider the existence of such leptophilic boosted 
 regions to be  a strong enough motivation for us to seek  compatibility of this
MSSM parameter space with recent cosmic ray observations.

In the following sections, we focus primarily on the dark matter mass of the electroweak scale.
We allow  the branching fraction $Br(\chi \chi \rightarrow \tau^+ \tau^-)$ to vary  up to $90$\% with $Br(\chi \chi \rightarrow  \tau^+ \tau^-)+BR(\chi \chi \rightarrow b \bar{b}) \approx 1$ as expected for the Higgs resonance regimes in the uplifted Higgs regions \cite{bog}.
This study should suffice for the purpose of pointing out the potential 
significance of previously missed MSSM parameter regions in view of the 
recent cosmic ray data. 

\section{Cosmic Ray Signals}
\label{cosray}

In this section we discuss positron, antiproton, and gamma ray signals in our scenario.
Henceforth for the density profile of the galactic halo we take the cored isothermal profile \cite{bah,core} 
to alleviate the stringent gamma ray constraints as we shall see, and we use the MED model 
\cite{don03,del} for the propagation parameters unless stated otherwise. 
We also use the force field approximation \cite{perko,glee} with the solar modulation parameter (Fisk potential) $\phi_F=0.6$~GV for definiteness for the charge independent solar modulation effects and 
do not consider the charge dependent solar modulation \cite{bal3,clem} whose realistic treatment is left for future work due to large uncertainties.

\subsection{Positron signals}
We attempt to explain
PAMELA $e^+/(e^++e^-)$ data as the signature of  dark matter annihilation.
However, we treat FERMI $e^++e^-$ data as  background; then the DM additional
contribution to $e^++e^-$ must be at most a small addition on top of this background.

Positrons  in the Galactic halo 
travel under the influence of the interstellar magnetic fields 
and lose energy via  inverse Compton scattering and synchrotron radiation.
Such effects (essentially random walk processes) 
can be treated via the steady-state diffusion equation for the 
number density per unit energy $N(E,{\bf x})$ \cite{text,ato,kob,hisa3}
\ba
\frac{\partial N}{\partial t}=0=\nabla \cdot [K(E,{\bf x})\cdot \nabla N]+\frac{\partial}{\partial E}
[b(E,{\bf x}) N]+Q(E,{\bf x})  \nonumber
\ea
where $K$ is the diffusion coefficient and $b=(1/\tau)\times E^2/(1GeV)$ is the energy loss rate 
(we take $\tau=10^{16}$ seconds) and $Q$ is the 
source term which is, in the dark matter annihilation scenarios, proportional to the dark matter number density squared and 
the thermally averaged annihilation cross section. 

\begin{figure}[htb!]
\begin{center}    
\epsfxsize = 0.48\textwidth
\epsffile{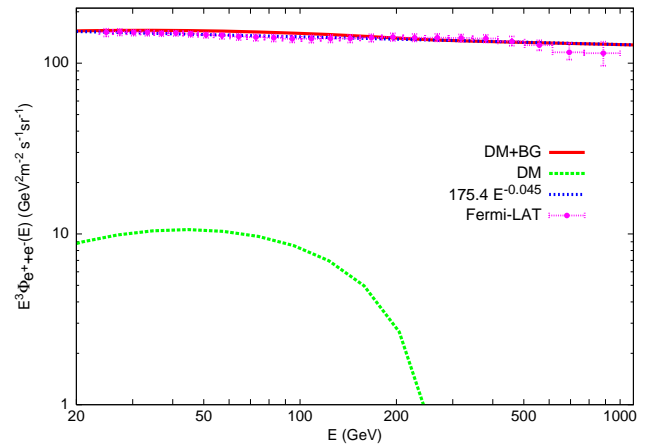}
\end{center}        
\caption{$e^++e^-$ flux. The FERMI data are shown, together with a power law fit to the data,
  $175.4({E}/{1GeV})^{-3.045} GeV^{-1}m^{-2}s^{-1}sr^{-1}$. The lower (green) curve indicates  the
   DM contribution  for $m_{\chi}=300$GeV, BF=110, BR($\chi \chi \rightarrow \tau^+ \tau^-$)=0.9 BR($\chi \chi \rightarrow b \bar{b}$)=0.1. The highest curve, which is the sum of the DM
   contribution plus the FERMI fit, is still within the systematic errors of the FERMI data.}
\label{esum1}
\end{figure}

First we discuss our treatment of the FERMI $e^++e^-$ data as background. A simple featureless power law, for instance the one in Ref. \cite{gra}
\ba
\Phi_{e^+}+\Phi_{e^-}=175.4\left(\frac{E}{1GeV}\right)^{-3.045} GeV^{-1}m^{-2}s^{-1}sr^{-1}
\nonumber
\ea
can fit the FERMI $e^++e^-$ data as shown in Figure \ref{esum1} (the statistical and systematic errors are added in quadrature). 
\begin{figure}
\begin{center}    
\epsfxsize = 0.48\textwidth
\epsffile{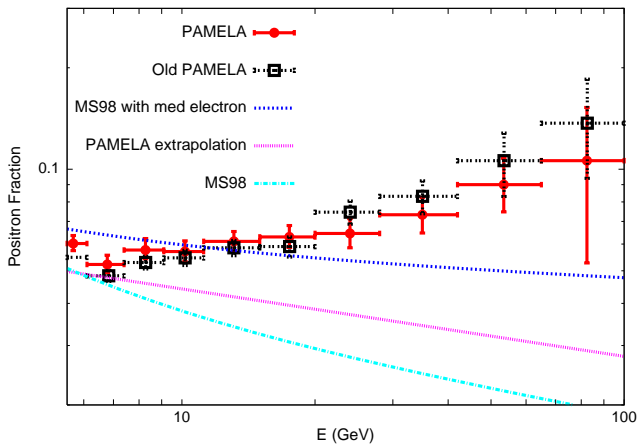}
\end{center}        
\caption{The uncertainties in the positron fraction backgrounds. The most recent PAMELA
data have a somewhat lower positron fraction than the earlier analysis. As our backgrounds,
we use the second curve from the bottom (the pink dotted line), which is an
extrapolation of the low-energy PAMELA data and lies in between the widely-used higher
Moskalenko and Strong (MS98) result and a fit with the spectral index $\gamma = 3.44$ (labeled med
electron). }
\label{posiBG}
\end{figure}
\begin{figure}
\begin{center}    
\epsfxsize = 0.48\textwidth
\epsffile{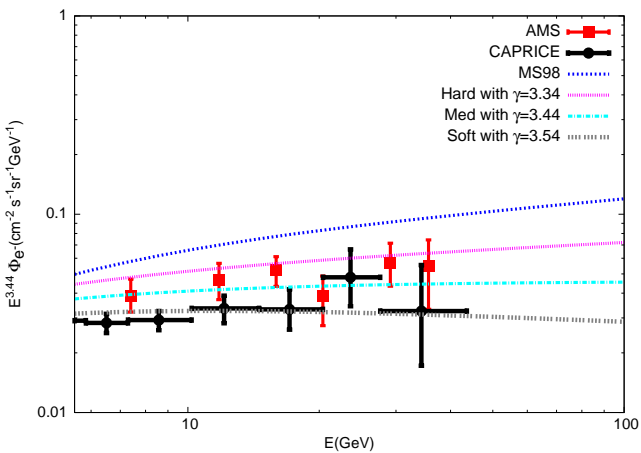}
\end{center}        
\caption{The electron flux background $\Phi_{e^-}$.}
\label{eminus}
\end{figure}
\begin{figure}
\begin{center}    
\epsfxsize = 0.48\textwidth
\epsffile{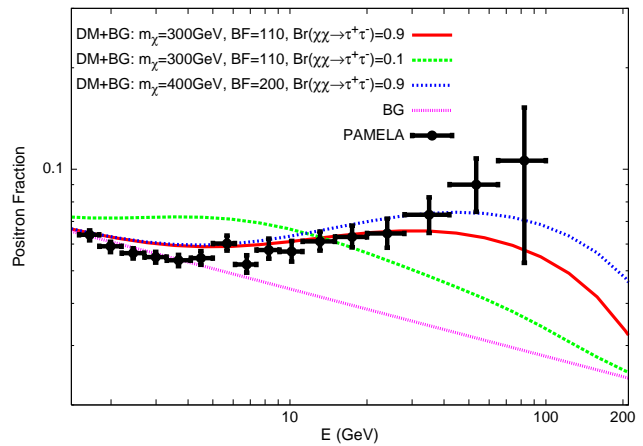}
\end{center}        
\caption{The positron fraction $e^+/(e^++e^-)$. The branching ratio of DM annihilation 
to taus is as labeled, where BR($\chi \chi \rightarrow \tau^+ \tau^-$)+BR($\chi \chi \rightarrow b \bar{b}$)=1. Boost Factors for the different curves are as labeled.}
\label{efrac1}
\end{figure}
We are in this section primarily interested in the dark matter mass of order a few hundred GeV and the boost factor $\sim {\mathcal O}(100)$ which can affect the lower energy part of FERMI data by order $\sim {\cal O}(10)$\% as illustarted in Figure \ref{esum1}. Such a 10\% shift in the spectrum can well be within the systematic errors \cite{gra}, and we do not expect, in the scenarios of our interest in this paper, the dark matter annihilations to affect significantly the current FERMI $e^++e^-$ spectrum.

Now we turn to PAMELA data.
In contrast to the relatively smooth FERMI $e^++e^-$ spectra, the bump-like rising 
behavior of the positron excess by PAMELA would be hard to  represent
 merely with  a simple power law, and we 
interpret this anomalous positron excess as the signal of  dark matter annihilations. 

For the astrophysical background of the positron fraction, we make a simple featureless power law fit for the PAMELA low energy data set 
\ba
\frac{\Phi_{e^+}}{\Phi_{e^+}+\Phi_{e^-}}=0.07 \left(\frac{E}{1GeV}\right)^{-0.2}
\ea
which was conservatively obtained from the lowest 6 energy bins out of the total 16 bins in 
the PAMELA data recently analyzed in Ref. \cite{adri2} without taking the charge-dependent solar modulation (whose uncertainty is quite large). 
 We extrapolate this fit
to higher energies and take it to be the $e^+/(e^++e^-)$ background.
This fitting formula gives
a background higher than the one conventionally used (as we discuss shortly). Including the higher energy bin data in 
order to get a fit makes the background even bigger.
A few comments on this positron fraction background are in order. Figure \ref{posiBG} 
illustrates the uncertainties in 
the positron fraction background estimations. Our background obtained by a simple 
extrapolation of PAMELA data is bigger than the conventionally used 
positron fraction background obtained by the electron and positron fluxes of Moskalenko \& Strong (MS98) \cite{mos,bal3}. 
The widely used MS98 fits can potentially underestimate the positron fraction partly because of 
the overestimation of the 
electron flux \cite{dela} as shown in Figure \ref{eminus}, which contrasts the MS98 fits
with the more recent electron flux data points from CAPRICE and AMS \cite{caprice94,ams01}. 
Figure \ref{eminus} also shows the electron flux parametrized by a power law with  spectral index  
$\gamma=3.44 \pm 0.1$ \cite{dela,casa}, which 
better represents the recent electron flux data and gives  a smaller $e^-$ amplitude than that used by Moskalenko \& Strong. The three indices $\gamma$ = 3.34, 3.44, and 3.54
are labeled "hard electron", "medium electron" and "soft electron" respectively in the figure.
The positron fraction background obtained by combining the electron flux with $\gamma=3.44$ 
(labeled med) together with  the positron flux estimation by Moskalenko \& Strong 
are also shown in Figure \ref{posiBG}. 
The current data of the positron flux itself still have too large error bars for  parametrization fitting 
to be feasible; however the forthcoming 
positron flux data such as those from AMS-02 \cite{ams02} would clarify these issues.
Figure \ref{posiBG} also shows the previous PAMELA analysis \cite{adri1} (denoted as old PAMELA) for reference. For definiteness, in the following, we use the 
most recent PAMELA analysis \cite{adri2} (denoted simply as PAMELA) and the simple 
PAMELA low energy data extrapolation for our positron fraction background.
Our choice of background seems reasonable because it  lies in  between
the conventional MS98 positron fraction background and the alternative background
obtained by using MS98 $e^+$ together with the "medium electron" fit to the currently available $e^-$ data
as can be seen in Fig \ref{posiBG}. 


Figure \ref{efrac1} shows the PAMELA positron fraction excess data together with the
predicted positron fraction from DM annihilation in our scenarios. 
One can see that the following parameter choice
is a good fit to data: the boost factor BF=110,  BR($\chi \chi \rightarrow \tau^+ \tau^-$)=0.9, and 
 BR($\chi \chi \rightarrow b \bar{b}$)=0.1.  For this parameter choice,
 the averaged $\chi^2$ for 10 highest energy bins is 1.0 and that for the 8 highest energy bins 
 (the choice of these bins is as conventionally chosen in the literature to avoid the effects of the charge dependent 
solar modulation) is 0.7. 
 Since this set of parameters is a good fit to PAMELA data, we take it 
  to be our {\it canonical case} for upcoming discussions of other cosmic ray signals.
Indeed this choice will illustrate
the existence of the MSSM parameter space compatible with the current cosmic ray data.  
We note that, had we chosen a different background fit
with a bigger amplitude such as the one shown in Figure \ref{posiBG}, a good match of
our predictions to the PAMELA data would be even easier  
because the required contribution from DM  would be smaller (e.g. the boost factor
could be smaller).  

In contrast, for the case with DM annihilation primarily to $b \bar b$,
the match to PAMELA data is quite poor. This is the typical situation in the usual MSSM. In the figure we illustrate the
  case with BR($\chi \chi \rightarrow \tau^+ \tau^-$)=0.1 and
  BR($\chi \chi \rightarrow b \bar{b}$)=0.9;  here the curves shift toward  lower energy because of 
  the softer positrons from $b$ compared with those from tau decays. 
  
  One can obtain a harder positron fraction spectrum by choosing a higher dark matter mass (the cutoff scale of the 
positron fraction spectrum shifts towards  higher energy for  kinematic reasons) and an example 
for  $m_{\chi}=400$GeV is shown for comparison. The higher dark matter mass decreases the 
overall amplitude due to the smaller number density, and hence we compensated for it with a 
bigger boost factor 200 leading to the averaged $\chi^2=1.0$ for 10 highest energy bins and
 $\chi^2=1.4$ for the 8 highest energy bins. 

Even though the exclusive 
parameter scanning for the uplifted Higgs regions in the MSSM parameter space is beyond the scope of this paper due to the nontrivial particle physics constraints (such as those from  flavor changing interactions), we can see that a wide range of  MSSM parameters 
in the uplifted Higgs region can be consistent with the current positron fraction data. 

Now we turn to constraints from other cosmic ray data for consistency. 

\subsection{Antiproton Flux}
The antiproton flux could be problematic because our scenario expects the $b \bar{b}$ 
annihilation channel to be typically of order $10$\% or more, and 
the non-negligible hadronic processes can overproduce  antiprotons in excess of
what is observed.
The antiproton flux is obtained by solving the steady diffusion equation \cite{don03,botti} analogously to that for
electron/positrons (except for 
the negligible energy loss term because  $m_p \gg m_e$)
\ba
\frac{\partial N}{\partial t}&=&0=K(T) \nabla^2 N -\frac{\partial }{\partial z}({\mbox sign} (z)N V_c)
\\ \nonumber
&& +Q({\bf x},T)-2h \delta(z)\Gamma_{ann}N
\ea
where $T$ is the antiproton kinetic energy, $V_c$ is a constant galactic convective wind (corresponding to 
a constant flow of magnetic irregularities) and $h=0.1$kpc is the 
height of our Galaxy approximated as a thin disk, $\Gamma_{ann}$ denotes the 
annihilation rate of antiproton and the interstellar proton. 
Even though there still exist large uncertainties in the background estimations of the 
antiproton-to-proton ratio, we present the following discussions using the secondary antiproton background estimated by the fitting formula \cite{bri} 
\ba
\log_{10}\phi_{\bar{p}}^{BG}=0.028 log ^4_{10} (T/GeV) -0.02 log^3_{10}(T/GeV) 
\nonumber \\  
-1.0 \log ^2_{10}(T/GeV)+0.07 \log _{10}(T/GeV)-1.64 \nonumber
\ea
and that for the primary protons via \cite{ptu}
\ba
\phi_p^{BG}(T)=\frac{890(T/GeV)^{-0.43}}{1. - 0.112(T/GeV)^{0.992} + 0.156(T/GeV)^{2.03}} \nonumber
\ea

\begin{figure}
\begin{center}    
\epsfxsize = 0.48\textwidth
\epsffile{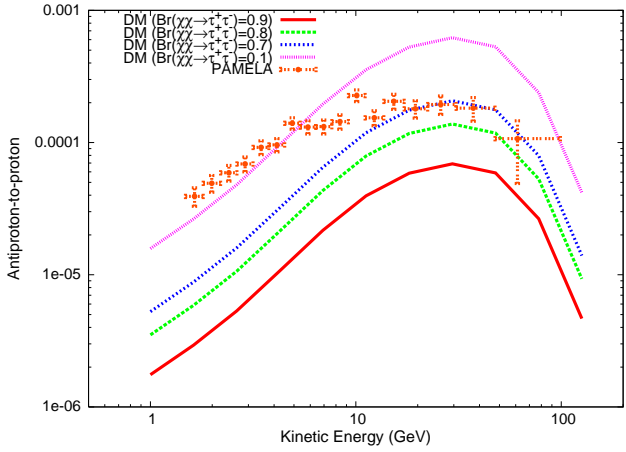}
\end{center}        
\caption{The antiproton-to-proton ratio ${\bar p}/p$ for different branching fractions with $m_{\chi}=300$GeV and BF=110. 
$Br(\chi \chi\rightarrow \tau^+ \tau^-)+Br(\chi \chi\rightarrow b \bar{b})=1$ in our model. In this figure,
 $Br(\chi \chi\rightarrow \tau^+ \tau^-)\sim 0.1$ 
is typical for the usual MSSM.  All the other curves, however, cannot be found in the usual MSSM but
are possible in the uplifted Higgs region discussed in this paper. }
\label{pbarnoBG}
\end{figure}

Figure \ref{pbarnoBG} shows the dark matter contributions to the antiproton-to-proton ratio for $m_{\chi}=300$ and BF=110, where we conservatively treated the PAMELA data \cite{pamelapbar} as the upper bound because of the large uncertainties in the background estimations. 
Figure \ref{pbarnoBG} clearly illustrates the 
advantage of the uplifted Higgs regions where the branching fractions into taus can be quite large compared with other conventionally 
studied MSSM parameter regions with a large $\tan \beta$. In fact, the conventional MSSM parameter space is typically not compatible with the lack of
 antiproton excess in the PAMELA. Indeed
 the cases with branching ratio into $b\bar{b}$ well over a few tens of percent are strongly
constrained.  Taking advantage of a small branching fraction to
$b\bar{b}$ of $\gtrsim {\cal O}(10)$\% in our dark matter annihilation scenarios, 
we can see that our scenarios can be compatible with the lack of the antiproton excess.

\begin{figure}
\begin{center}    
\epsfxsize = 0.48\textwidth
\epsffile{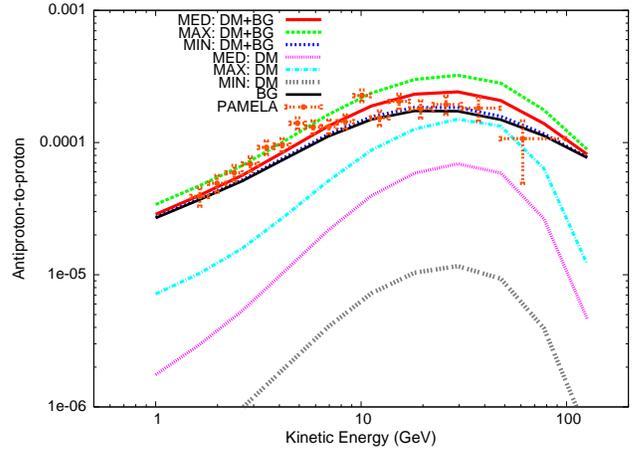}
\end{center}        
\caption{The antiproton-to-proton ratio ${\bar p}/p$ including the background contributions for different propagation parameters with $m_{\chi}=300$GeV, BF=110, $Br(\chi \chi\rightarrow \tau^+ \tau^-)=0.9$ and $Br(\chi \chi\rightarrow b \bar{b})=0.1$. }
\label{pbar}
\end{figure}

For illustration purposes, we also 
added the background contributions to the antiproton-to-proton ratio in Figure \ref{pbar} which certainly tighten the constraints.

For a reference, we also showed in Fig. \ref{pbar}, in addition to 
MED model we have been using throughout the paper, the MIN and MAX models to indicate the further uncertainties arising from the propagation parameters \cite{don03,del}. MAX models would have rather stringent constraints from PAMELA $\bar{p}/p$ data while the antiproton productions for MED and MIN parameters can still be within the observed bound. Despite those uncertainties, one can see that the our MSSM scenarios with the ${\cal O}(10)$\% $b\bar{b}$ channel can be compatible with the currently available ${\bar p}/p$ data while producing the positron excess.



\begin{figure}
\begin{center}    
\epsfxsize = 0.48\textwidth
\epsffile{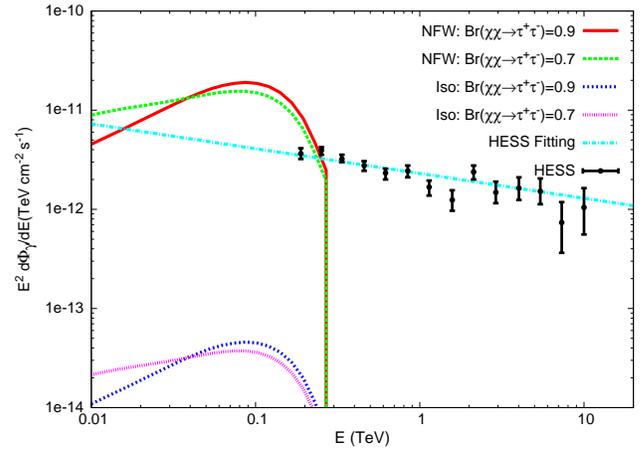}
\end{center}        
\caption{The gamma ray fluxes from the Galactic Center due to DM annihilation
 for the solid angle $\Delta \Omega=10^{-5}$sr. $m_{\chi}$=300GeV, BF=110, BR($\chi \chi 
 \rightarrow \tau^+ \tau^-) +$BR$(\chi \chi \rightarrow b \bar{b})$=1. We also show, for reference, 
 the HESS data along with its fitting function. NFW and Iso refer to NFW and isothermal halo profiles.}
\label{hess04}
\end{figure}
\begin{figure}
\begin{center}    
\epsfxsize = 0.48\textwidth
\epsffile{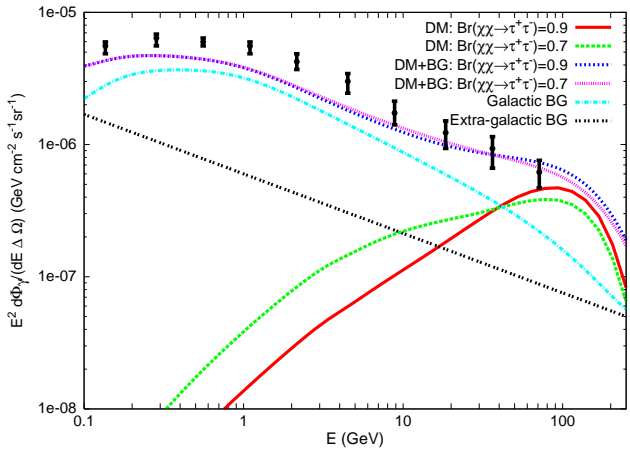 }
\end{center}        
\caption{The gamma ray fluxes for $10^{\circ}<|b|<20^{\circ},0^{\circ}<l<360^{\circ}$. $m_{\chi}=300$GeV, BF=110, BR($\chi \chi \rightarrow \tau^+ \tau^-) +$BR$(\chi \chi \rightarrow b \bar{b})$=1. Data are from FERMI and the isothermal profile is used.  The background (BG) that
is added to the DM signals is the sum of Galactic and Extragalactic backgrounds, each
of which is plotted here.}
\label{fermigamma2}
\end{figure}

\subsection{Gamma Ray Constraints}
 The $\gamma$-ray data can also potentially exclude some of uplifted Higgs regions because our scenarios expect to give a significant $\gamma$-ray flux 
from $\tau$ decays \cite{gamma2,gian,serp,gammabd,pap,cirel,kev2}. The $\gamma$-rays, being chargeless, travel undeflected, and the flux from DM  annihilation is given by
\ba
\frac{d \Phi_{\gamma}}{dE_{\gamma}}=
\frac{1}{2}
\frac{r_{\odot} \rho_{\odot} ^2 \langle \sigma v \rangle}
{4 \pi m_{\chi}^2}
\frac{dN_{\gamma}}{dE_{\gamma}}
\bar{J}(\Delta \Omega)
\Delta \Omega
\ea
with 
\ba
\bar{J}(\Delta \Omega)
&\equiv&
\frac{1}{\Delta \Omega}
\int_{\Delta \Omega} J(\psi)d\Omega
\ea  
and
\ba
J(\psi)=\frac{1}{r_{\odot} \rho_{\odot} ^2 }
\int_0^{\infty} \rho^2(r) dl(\psi)
\ea
where $\psi$ is the angle between the line of sight and the direction of the galactic center, 
$r^2=l^2+\rho_{\odot} ^2-2l\rho_{\odot} \cos \psi$ and $\Delta \Omega=2\pi(1-\cos \psi)$ is the observed region of the sky. 
 Stringent $\gamma$-ray constraints on our DM annihilation 
scenarios with the dominant tau channel arise from the prompt $\gamma$-rays
due to the fragmentation of the annihilation products.

The previous gamma ray data even before FERMI such as 
HESS \cite{hess1} already offer  tight constraints. 
Figure \ref{hess04} illustrates
the HESS data from the Galactic Center, its fitting function 
$E^2 d \Phi_{\gamma}/dE= 2.3 \times 10^{-12} (E/TeV)^{-2.25} [TeV^{-1} cm^{-2} s^{-1} ]$, as well as a number of predictions from DM annihilation.  We have
plotted the gamma ray flux from the Galactic Center (GC) for our canonical example due to the hadronic process calculated by DarkSUSY \cite{dark} where Pythia \cite{pythia} is implemented. For comparison, we also plot the $\gamma$-rays for a smaller tau 
channel branching ratio, which makes the spectra softer as expected. 

 The DM predictions towards the GC within the resolved direction
  can be strongly dependent on the halo density profiles. The Navarro Frenk White (NFW)
   profile \cite{nfw}, for instance,  has a density that is strongly peaked towards the GC
   and leads to $\gamma$-ray fluxes  in excess of observations.  The isothermal
   profile, which has a central core, predicts fewer $\gamma$-rays and is in better
   agreement with data. For this reason we have been using the isothermal profile throughout the paper.

An unprecedented plethora of all-sky gamma ray data from FERMI offers
 further constraints on  dark matter annihilation scenarios. 

The prompt gamma ray contributions along with the the preliminary FERMI data in the mid latitude $10<b<20$ \cite{port3} 
extending the energy range published in Ref \cite{port1} are shown in Fig. \ref{fermigamma2}. This can give an upper bound on the gamma ray flux from 
the dark matter annihilations in our scenarios.

To be more restrictive, we also show, for illustration,
 the background contributions (which however  have large uncertainties as well as 
 contamination from  point sources). For a simple estimation of the background, we added the Galactic background contributions (the sum of contributions from inverse Compton scattering, Bremsstrahlung and $\pi^0$) of the conventional GALPROP model in Ref. \cite{stro4} and the isotropic diffuse extragalactic $\gamma$-ray background from FERMI parametrized by the power law $\Phi^{EG}_{\gamma}\propto {E}^{-2.41}$ \cite{acker}. We can see, despite the the uncertainties in the background estimation, that the 
dark matter annihilation scenarios in the MSSM with the electroweak scale mass 
can be still compatible with the current $\gamma$-ray data. 

It will be of great interest to see what insights into supersymmetric model
building can be obtained from upcoming   cosmic and $\gamma$-ray data with  reduced systematic errors and 
improved background estimation, particularly when combined with upcoming data from the Large Hadron Collider.


\section{Conclusion/Discussion}
\label{conc}
We have illustrated that  dark matter annihilation scenarios in the MSSM can be viable 
candidates to explain all current cosmic ray observations without necessarily 
introducing any additional degrees of freedom.   In particular, we have studied pseudoscalar
Higgs s-channel resonance in the uplifted Higgs region to obtain boosted leptophilic
annihilation cross sections which can explain PAMELA data while not in conflict
with any other data sets.

We here mention several directions for possible future improvement of our analysis.
The positron fraction constraints mainly come from the highest energy bins of the data with large error bars and it should improve by the forthcoming 
data from PAMELA and AMS-02 with better controls of the systematic/statistical errors. A proper treatment of the charge dependent solar modulation effects (which 
we did not consider) could account for discrepancies in the low energy data ($E\lesssim 10$GeV) among different experiments. In addition to the positron fraction which also suffer from the 
electron background estimation uncertainties, the absolute flux of positrons such as those from AMS-02 would give more definite probe of the underlying physics.


Our studies assumed the same value of the boost factor for all species of cosmic rays.
However, this is not necessarily the case. 
In particular, the antiprotons we observe originate from a large region of the halo 
in contrast to the positrons which are produced locally (say within a few kpc).
Hence, in principle, the antiproton flux could be less boosted than 
the positron flux, for instance if the positrons are boosted partially by the local clumpiness  \cite{boost2, die2,kuh,alb,brun,mark1}. Such 
relaxation of the antiproton constraints could be important for the 
annihilation scenarios within the framework of the MSSM, most of which 
are excluded due to antiproton overproduction. 

We presented our discussion primarily assuming an isothermal 
profile for the galactic halo.  In fact a more cuspy profile towards the center,
such as an NFW profile, was shown to be severely constrained \cite{cirel,pap,kev2}. Currently
there is a great deal of uncertainty regarding the density distribution
in the inner regions of galaxies, though there exist observations of flattened cores \cite{salucci} in spiral galaxies. More precise 
observations as well as detailed simulations including the proper treatment of the gas physics
will be required for more realistic modeling of the halo profiles.

Further studies of $\gamma$-ray constraints beyond the analysis of this paper
would be warranted.
We refer the reader to, for instance, Ref. \cite{pap,cirel,kev2,regis} for gamma ray constraints 
from additional regions of the sky and halo profiles. We focused on 
gamma rays from the hadronic processes which are characteristic of
DM annihilation scenarios 
in which $\tau$ channel dominates;  the consideration of additional 
effects such as the inverse Compton scattering including those 
to the extragalactic gamma rays \cite{stepha1,danbbb,kawa} could 
give additional constraints depending on the 
parameter ranges of interest. 

More stringent constraints could however
 come from the particle physics  rather than from the astrophysics once we have a concrete particle physics
model.   For instance, for the uplifted Higgs scenario, the tuning of the heavy Higgs decay width 
$\Gamma_A/m_A \lesssim {\cal O}(10^{-3}$) implies 
$y_{\tau}\lesssim {\cal O}(10^{-1})$, which needs to be checked with  flavor physics constraints. We also could in principle 
consider a large $y_{\mu}$ in an analogous manner to a large $y_{\tau}$ by adjusting the 
slepton, squark masses and the phase of the gluon mass \cite{bog}.  If $\mu$ final states could be significant by such 
(possibly fine-tuned) adjustments, then the $\gamma$-ray constraints could be relaxed 
relative to the $\tau$ dominant scenarios considered in this paper 
\cite{pap,cirel,mea,eds} even though here again the particle physics constraints (e.g. the flavor changing 
neutral currents) would need to be carefully checked.

In the vast MSSM parameter space, it is worth searching for other regions previously missed
that could explain all the cosmic ray data.  The uplifted Higgs region itself deserves further
study.  Even though our cosmic ray analysis is based on the properties of the 
uplifted Higgs regions, we kept our analysis fairly general so that  a similar study
 could be applicable to any other parameter regions with similar properties.
 The unprecedented wealth of data expected from upcoming astrophysical and terrestrial 
experiments (e.g. AMS-02 and LHC) could well provide further motivation towards
 a fuller exploration of the MSSM parameter space in the coming years.

\section*{Acknowledgments}
We thank B. Dobrescu, D. Feldman and G. Kane for  useful discussions. KK especially thanks B. Dobrescu for his encouragement to pursue the astrophysical 
signatures of the uplifted Higgs regions. This work in part was supported by the Michigan Center for Theoretical Physics and the DOE at the University of Michigan and by NSF award PHY-0456825 and NASA award NNX09AT09AT70G at the University of Utah.

\end{document}